\documentclass[prl,aps,preprintnumbers,superscriptaddress,twocolumn,showpacs]{revtex4}
\usepackage{graphicx,xspace,amsmath,amssymb,amsfonts}
\newcommand{\ket}[1]{\ensuremath{| #1 \rangle}}
\newcommand{\bra}[1]{\ensuremath{\langle #1 |}}
\newcommand{\Sum}[2]{\ensuremath{\sum_{#1}^{#2}}}
\newcommand{\om}[1]{\ensuremath{\omega^{#1}}}

\newcommand{\compi}{\ensuremath{\mathrm{i}}}
\newcommand{\pr}[1]{\ensuremath{\mathrm{Prob\!}\left(\mbox{#1}\right)}}

\newcommand{\ch}[2]{\ensuremath{\left(
\begin{array}{c}
  #1 \\
  #2 \\
\end{array}
\right)}}
\begin{document}

\title{Quantum tomographic cryptography with Bell diagonal states:
  non-equivalence of classical and quantum distillation protocols}%

\author{Dagomir Kaszlikowski}
\affiliation{Department of Physics, National University of Singapore,
Singapore 117\,542, Singapore}

\author{Jenn Yang, Lim}
\affiliation{Department of Physics, National University of Singapore,
Singapore 117\,542, Singapore}

\author{D. K. L. Oi}
\affiliation{Centre for Quantum Computation, Department of Applied Mathematics
  and Theoretical Physics, University of Cambridge, Cambridge CB3 0WA, U.K.}

\author{Frederick H.\ Willeboordse}
\affiliation{Department of Physics, National University of Singapore,
Singapore 117\,542, Singapore}

\author{Ajay Gopinathan}
\affiliation{National Institute of Education,
Nanyang Technological University, Singapore 639\,798, Singapore}
\affiliation{Department of Physics, National University of Singapore,
Singapore 117\,542, Singapore}

\author{L. C. Kwek}
\affiliation{National Institute of Education,
Nanyang Technological University, Singapore 639\,798, Singapore}
\affiliation{Department of Physics, National University of Singapore,
Singapore 117\,542, Singapore}


\begin{abstract}
  We present a generalized tomographic quantum key distribution protocol in
  which the two parties share a Bell diagonal mixed state of two qubits. We
  show that if an eavesdropper performs a coherent measurement on many quantum
  ancilla states simultaneously, classical methods of secure key distillation
  are less effective than quantum entanglement distillation protocols. We also
  show that certain Bell diagonal states are resistant to any attempt of
  incoherent eavesdropping.
\end{abstract}
\maketitle

\section{Introduction}

The security of quantum key distribution (QKD)~\cite{BB84, Ekert91} is an
important consequence of the application of the laws of physics to information
and communication theory. A one-time pad provides perfect cryptographic
security for sending messages between two parties but relies on being able to
distribute a shared secret key~\cite{vernam,walsh}. Classically, it is
impossible to amplify a set of shared randomness, but quantum mechanics allows
this to be done by the transmission of quantum states~\cite{bennett92}. The
full power of quantum cryptography rests on the ability to place upper bounds
on the knowledge of a potential eavesdropper (Eve) about the distributed key
shared by the legitimate parties (Alice and Bob). In this paper we present a
generalization of the so-called tomographic quantum key distribution
protocol~\cite{Cherng}. We consider the situation where Alice and Bob use qubits in a maximally entangled state distributed by a central source. The qubits undergo a quantum channel that converts the state to a Bell diagonal mixed state.

We analyze the security of this protocol under two broad scenarios. In the first
scenario, Alice and Bob agree on a cryptographic key if the correlations
between their measurement results are stronger than any possible correlations
between one of them and Eve, under the assumption
that Eve has full control over the source of entangled qubits but she can only
perform incoherent measurements. The tomographic element of the protocol
allows Alice and Bob to compute the maximal strength of correlations between
Eve and any one of them. The Csisz{\'a}r-K{\"o}rner~\cite{CK} theorem then
guarantees that if the correlations between Alice and Bob are stronger than
those between Eve and either of them, a secure key can be established through
\emph{one-way} error correcting codes.

In the second scenario, we examine the situation when Eve's
correlations are initially stronger than Alice and Bob's. It was
shown that in some cases it is still possible to obtain a secure
key \cite{Maurer}. The idea is that by means of two-way
communication Alice and Bob can strengthen their correlations with
respect to Eve's so that the CK theorem can be applied again. This
procedure is called \emph{advantage distillation} (AD).

There are two possible strategies for Eve within the second scenario:
incoherent and coherent measurements. The first case was examined
in~\cite{ACIN} where it was shown that advantage distillation is possible as
long as the two-qubit state shared by Alice and Bob is entangled. We re-derive
this result using different reasoning than the one presented in~\cite{ACIN}.

In the second case, we show that the above result no longer holds in the case
of coherent measurements by Eve. Indeed, if the qubits are affected by too many
errors (caused by Eve's actions), advantage distillation fails despite Alice and
Bob still sharing an entangled state.  In such cases the only way for Alice and
Bob to obtain a secure key is to revert to quantum entanglement distillation.

\section{Tomographic QKD}

In a tomographic QKD scheme, a central source distributes
entangled qubits to Alice and Bob. They independently and randomly
choose to measure three tomographically complete observables
$\sigma_x$, $\sigma_y$ and $\sigma_z$ (Pauli matrices) on each
qubit. At the end of the transmission, they publicly announce
their choice of observables for each qubit pair. They then proceed
to divide their measurement results according to those for which
their measurement bases match, and those for which their
measurement bases do not match. Exchanging a subset of their
measurements allows Alice and Bob to tomographically reconstruct
the density operator of the two-qubit state they share.

Ideally, in the absence of noise in the source or channels, they
expect to receive the maximally entangled state
\begin{equation}\label{ideal}
\ket{\psi_{ideal}}=\frac{1}{\sqrt{2}}\left(\ket{z_0,z_0}+\ket{z_1,z_1}\right),
\end{equation}
where $\ket{z_{k}}$ is the eigenstate of $\sigma_z$ with the
eigenvalue $(-1)^{k}$, and Alice (Bob) possesses the left (right)
qubit.
The results for matching bases can then be used to generate a
cryptographic key as they are either perfectly correlated ( for
$x$ and $z$ bases) or anti-correlated (for $y$ basis).

However, Alice and Bob cannot realistically expect to obtain the maximally
entangled state Eq.~(\ref{ideal}) because either the source is not ideal, the
channel conveying the qubits is noisy, or there is an eavesdropper tampering
with the source. For security analysis, we assume that Eve has total control
over the source and that all the errors are caused by her when she tries to
extract information about the key.

To constrain Eve's information, Alice and Bob use part of their measurements to
perform full tomography on the state distributed by the source. The protocol
we consider here is such that Alice and Bob agree to communicate if and only
if they see the Bell diagonal state
\begin{eqnarray}\label{theform}
\varrho_{AB} &=& \Sum{a,b=0}{1}p_{ab}\ket{z_{ab}}\bra{z_{ab}},
\end{eqnarray}
where
\begin{eqnarray}
\ket{z_{ab}} &=& \frac{1}{\sqrt{2}}\Sum{k=0}{1}\om{kb} \ket{z_k,
z_{k+a}}
\end{eqnarray}
and $\Sum{a,b=0}{1}p_{ab}=1, \om{}=-1$. Following the nomenclature
of~\cite{Hashing} we call $a$ the amplitude bit and $b$ the phase
bit. Here, we assume that $p_{00} > \frac{1}{2}$ \cite{p00}.

The above state can be obtained from the maximally entangled state
Eq.~(\ref{ideal}) assuming that the travelling qubits undergo bit
and phase flips. The so-called Werner state, i.e., the maximally
entangled state with white noise, is a special case where
$p_{01}=p_{10}=p_{11}$. Therefore, the protocol presented here is
more general than the one studied in~\cite{cadIN, Cherng} where
only Werner states were considered.

As Alice and Bob perform their measurements in the three bases
$x,y$ and $z$, it is convenient to express the state
$\varrho_{AB}$ in the $x$ and $y$ bases. This can easily be done
using the transformation rules on the Bell states,
\begin{equation}
\ket{z_{ab}}=\om{ab}\ket{x_{ba}}=(-\compi)^a\om{ab}\ket{y_{a+b+1\;
a}}.
\end{equation}
Writing out (\ref{theform}) in the other bases, we have
\begin{equation}
\varrho_{AB}=\Sum{a,b=0}{1}p_{ba}\ket{x_{ab}}\bra{x_{ab}}=
\Sum{a,b=0}{1}p_{b\;a+b+1}\ket{y_{ab}}\bra{y_{ab}}.
\end{equation}
We can then compute the probability of Alice and Bob obtaining
correlated results conditional on a particular choice of basis:
\begin{eqnarray}
\pr{correlation$|$$x$ basis} &=& p_{00}+p_{11}  \nonumber\\
\pr{correlation$|$$y$ basis} &=& p_{01}+p_{10}  \nonumber\\
\pr{correlation$|$$z$ basis} &=& p_{00}+p_{11},
\end{eqnarray}
and also the probability of getting anti-correlated results:
\begin{eqnarray}
\pr{anti-correlation$|$$x$ basis} &=& p_{01}+p_{10}  \nonumber\\
\pr{anti-correlation$|$$y$ basis} &=& p_{00}+p_{11}  \nonumber\\
\pr{anti-correlation$|$$z$ basis} &=& p_{01}+p_{10}.
\end{eqnarray}
Since $p_{00}>\frac{1}{2}$, Alice and Bob are more likely to
obtain correlated results when they measure in the $x$ and $z$
bases, and anti-correlated results in the $y$ basis; Alice and Bob
will thus make use of correlation to generate their key when they
measure in the $x$ and $z$ bases, and anti-correlation to generate
their key when in the $y$ basis.

\section{Eavesdropping}

In order to obtain as much information as possible about the key generated by
Alice and Bob, Eve entangles their qubits with ancilla states $\ket{e_{ab}}$
in her possession. The best she can do is to prepare the following tripartite
pure state
\begin{equation}
\ket{\psi_{ABE}} =\sum_{a,b=0}^{1} \sqrt{p_{ab}} |z_{ab}\rangle
|e_{ab}\rangle, \label{eq:eavestate}
\end{equation}
where $\langle e_{ab}|e_{cd}\rangle = \delta_{a,c}\delta_{b,d}$.
Tracing out Eve gives the mixed state Eq.~(\ref{theform}) that
Alice and Bob measures, and this purification is the most general
one as far as incoherent attacks are concerned.

Eve's purifications, when expressed in different bases, read
\begin{eqnarray}\label{evezstate1}
\ket{\psi_{ABE}}&=& \frac{1}{\sqrt{2}}
\Sum{k,a=0}{1}\ket{z_k,z_{k+a}}\left(\Sum{b=0}{1}\sqrt{p_{ab}}\om{kb}\ket{e_{ab}}\right)\nonumber\\
&=&
\frac{1}{\sqrt{2}}\Sum{k,a=0}{1}\ket{x_k,x_{k+a}}\left(\Sum{b=0}{1}\sqrt{p_{ba}}\om{kb}\om{ab}\ket{e_{ba}}\right)
\nonumber\\
&=& \frac{1}{\sqrt{2}}\Sum{k,a=0}{1}\ket{y_k,y_{k+a}} \nonumber\\
&& \;\; \left(\Sum{b=0}{1}\sqrt{p_{b\;a+b+1}}(-\compi)^b\om{kb}\om{b(a+b+1)}\ket{e_{b\; a+b+1}}\right),  \nonumber\\
\end{eqnarray}
We can express Eq.~(\ref{evezstate1}) more conveniently as
\begin{eqnarray}
|\psi_{ABE}\rangle&=&\Sum{k,a=0}{1}\sqrt{\frac{p_a}{2}}\ket{z_k,z_{k+a}}\ket{f_{ka}^{z}}
\nonumber\\
&=&
\Sum{k,a=0}{1}\sqrt{\frac{q_a}{2}}\ket{x_k,x_{k+a}}\ket{f_{ka}^{x}}
\nonumber\\
&=&
\Sum{k,a=0}{1}\sqrt{\frac{r_a}{2}}\ket{y_k,y_{k+a}}\ket{f_{ka}^{y}},
\end{eqnarray}
where
\begin{eqnarray}
p_a &=& \Sum{b=0}{1}p_{ab}  \nonumber\\
q_a &=& \Sum{b=0}{1}p_{ba} \nonumber\\
r_a &=& \Sum{b=0}{1} p_{b \; a+b+1}
\end{eqnarray}
and the normalized kets
\begin{eqnarray}
\ket{f_{ka}^{z}} &=&
\frac{1}{\sqrt{p_a}}\Sum{b=0}{1}\sqrt{p_{ab}}\om{kb}\ket{e_{ab}}
\nonumber\\
\ket{f_{ka}^{x}} &=&
\frac{1}{\sqrt{q_a}}\Sum{b=0}{1}\sqrt{p_{ba}}\om{kb}\om{ab}\ket{e_{ba}}
\nonumber\\
\ket{f_{ka}^{y}} &=& \frac{1}{\sqrt{r_a}}\Sum{b=0}{1}\sqrt{p_{b\;
a+b+1}}(-\compi)^b\om{kb}\om{b(a+b+1)}\ket{e_{b\; a+b+1}}
\nonumber\\
\end{eqnarray}
are such that their inner products are given by
\begin{eqnarray}
\langle f_{0a}^{z}|f_{1a}^{z} \rangle &=&
\frac{p_{a0}-p_{a1}}{p_{a0}+p_{a1}}\equiv \lambda^{z}_a  \nonumber\\
\langle f_{0a}^{x}|f_{1a}^{x} \rangle &=&
\frac{p_{0a}-p_{1a}}{p_{0a}+p_{1a}} \equiv\lambda^{x}_a \nonumber\\
\langle f_{0a}^{y}|f_{1a}^{y} \rangle &=&
\frac{p_{0\;a+1}-p_{1a}}{p_{0\;a+1}+p_{1a}} \equiv\lambda^{y}_a.
\end{eqnarray}
The ancillas with different $a$'s are orthogonal.

Eve's eavesdropping strategy proceeds as follows. After Alice and
Bob announce their measurement bases, Eve knows on which pairs of
qubits they measured the same observables and that her ancilla is
a mixture of four possible states.  Formally this can be viewed as
a transmission of information from Alice and Bob to Eve encoded in
the quantum state of Eve's ancilla. To find the optimal
eavesdropping strategy, she has to maximize this information
transfer by a choice of a suitable generalized measurement known
as a Positive Operator Value Measure (POVM).  For example, if
Alice and Bob measured in the $x$ basis, Eve will obtain the
following mixed state of her ancilla,
\begin{equation}
\varrho_E^x=\Sum{k,a=0}{1} \frac{q_a}{2} |f^{x}_{ka}\rangle\langle
f^{x}_{ka}|.
\end{equation}
This is equivalent to Alice and Bob ``communicating'' to Eve that
they measured $\{00,01,10,11\}$ by sending her the quantum states
$\{\ket{f^{x}_{00}},\ket{f^{x}_{01}},\ket{f^{x}_{11}},
\ket{f^{x}_{10}}\}$ with prior probabilities
$\{\frac{q_0}{2},\frac{q_1}{2},\frac{q_1}{2},\frac{q_0}{2}\}$
respectively. Eve has to find the optimal measurement that will
extract from the transmission as much information as possible,
called the \emph{accessible information}. Note that this is not
equivalent to finding a measurement that minimizes the error of
distinguishing between these states~\cite{SHOR}.

\section{Incoherent Attack}

We first assume that Eve carries out an \emph{incoherent} attack
in which she performs measurements on her ancillas one at a time.
In contrast, in a coherent attack, she would measure joint
observables of more than one ancilla, or construct her initial
state Eq.~(\ref{eq:eavestate}) so that more than one pair of
qubits were entangled with each ancilla.

The ancilla states for each basis can be divided into two groups. The first
group corresponds to $a=0$ and refers to the case when Alice and Bob obtain
correlated results. The second group corresponds to the case $a=1$ and refers
to the case when Alice and Bob obtain anti-correlated results.

For example, if Alice and Bob both measure in the $y$ basis, Eve
will have the state
\begin{equation}
\varrho_E^y=\Sum{k,a=0}{1} \frac{r_a}{2} |f^{y}_{ka}\rangle\langle
f^{y}_{ka}|.
\end{equation}
The first group $a=0$ occurs with probability $r_0$ and the second
group $a=1$ occurs with probability $r_1$. Similarly, if they
measure in the $x$ ($z$) basis, the first group occurs with
probability $q_0$ ($p_0$) while the second group occurs with
probability $q_1$ ($p_1$). The ancillas in the first group
$\ket{f_{k0}^{m}}$ ($m=x,y,z$) are orthogonal to those in the
second group $\ket{f_{k1}^{m}}$.

For the purpose of applying the Csisz{\'a}r-K{\"o}rner theorem, we
need only to compute the mutual information between Eve and Bob
and compare this with the mutual information between Alice and
Bob; Eve would have to optimize her measurements on her ancilla so
that it maximizes the information she gains about Bob's
measurement results.

Let us now present the POVM measurement that maximizes the
information transferred by Bob to Eve.
In the first step, Eve sorts the mixture of the ancillas into two
sub-ensembles according to the index $a$. This can easily be done
using a projective measurement. This sorting is an auxiliary step
as, at this stage, she does not gain any more information about
the result of Bob's measurement. After that, depending on the
outcome of the projection ($a=0$ or $a=1$), Eve has an
equiprobable mixture of two non-orthogonal ancilla states each
corresponding to Alice and Bob's result. For example, if the
chosen measurement basis was the $z$ basis, Eve will receive the
mixed state
\begin{eqnarray}
\varrho^z_E &=&
\Sum{k,a=0}{1}\frac{p_a}{2}\ket{f^z_{ka}}\bra{f^z_{ka}}.
\end{eqnarray}
Projecting into either the $a=0$ or $a=1$ orthogonal subspaces
(depending on Alice and Bob's measurement outcomes), she will
obtain one of the two equiprobable mixtures of non-orthogonal
ancilla states
\begin{eqnarray}
\varrho_0^z = \frac{1}{2} \ket{f^z_{00}}\bra{f^z_{00}}+\frac{1}{2}
\ket{f^z_{10}}\bra{f^z_{10}} &\mbox{if Alice and Bob obtained}&
\nonumber\\
&\mbox{correlated results, $a=0$;}&  \nonumber\\
\varrho_1^z = \frac{1}{2} \ket{f^z_{01}}\bra{f^z_{01}}+\frac{1}{2}
\ket{f^z_{11}}\bra{f^z_{11}} &\mbox{if Alice and Bob obtained}&
\nonumber\\
&\mbox{anti-correlated results, $a=1$.}&  \nonumber\\
\end{eqnarray}

Next, she applies the measurement that maximizes the accessible
information encoded in the mixture of the two ancilla states given
by the outcome of her projective measurement. In the case of two
equally likely states, this optimum measurement is given by the
so-called {\it square-root measurement}~\cite{sqm1, sqm2}. The
outcome probabilities of the square-root measurement are
\begin{eqnarray}
\eta_a^{x} &=& \frac{1}{2}\left(1+\sqrt{1-(\lambda_a^{x})^2}\right)\nonumber\\
\eta_a^{y} &=& \frac{1}{2}\left(1+\sqrt{1-(\lambda_a^{y})^2}\right)\nonumber\\
\eta_a^{z} &=&
\frac{1}{2}\left(1+\sqrt{1-(\lambda_a^{z})^2}\right),
\end{eqnarray}
where $\eta_a^{m}$ is the probability of correctly inferring a
given ancilla state in the $m$ basis ($m=x,y,z$). The index $a$
refers to the correlation/anti-correlation subspace in which the
ancilla lies.

It is straightforward to compute the mutual information between
Bob and Eve:
\begin{eqnarray}
I_{BE} &=&
\frac{1}{3}I_{BE}^{x}+\frac{1}{3}I_{BE}^{y}+\frac{1}{3}I_{BE}^{z},
\end{eqnarray}
where $I_{BE}^{m}$ is the mutual information when Alice and Bob
measure in the same basis $m$. We have
\begin{eqnarray}
I_{BE}^{x} &=& q_0 \left(1-H(\eta_0^{x})\right)+q_1
\left(1-H(\eta_1^{x})\right)  \nonumber\\
I_{BE}^{y} &=& r_0 \left(1-H(\eta_0^{y})\right)+r_1
\left(1-H(\eta_1^{y})\right)  \nonumber\\
I_{BE}^{z} &=& p_0 \left(1-H(\eta_0^{z})\right)+p_1
\left(1-H(\eta_1^{z})\right).
\end{eqnarray}
Here, $H(\eta_a^{m}) =
-\eta_a^{m}\log_2{\eta_a^{m}}-(1-\eta_a^{m})\log_2{(1-\eta_a^{m})}$
is the \emph{binary entropy} of the respective probability
distributions. Also, the mutual information between Alice and Bob
is given by
\begin{eqnarray}
I_{AB} &=& 1-\frac{1}{3}\left(H(p_0)+H(q_0)+H(r_0)\right).
\end{eqnarray}

We are interested in the conditions for which our protocol is secure against Eve's incoherent eavesdropping attack.
Now, even if Eve obtains some information about the transmitted key through her incoherent measurement,
Alice and Bob can still obtain a secure key with a few additional
steps. According to the Csisz{\'a}r-K{\"o}rner (CK) theorem, a
secure key can be generated from a raw key sequence by means of a
suitably chosen error-correcting code and classical one-way
communication between Alice and Bob if the mutual information
between Alice and Bob exceeds that between Eve and either one of
them (the CK regime). For the protocol considered, the mutual
information between Alice and Eve, and Bob and Eve, are the same
so that security is assured as long as
\begin{eqnarray}\label{MI}
I_{AB} &>& I_{BE}.
\end{eqnarray}

\section{Quantum Entanglement Distillation}

If there is too much noise in the two-qubit state, the CK theorem is
not immediately applicable. Instead, Alice and Bob need to either select a
subsequence of their bit values in a systematic way or pre-process their
two-qubit state before measuring, so that the CK theorem is applicable once
more. One method of doing this is \emph{quantum entanglement distillation} (QED), a quantum
procedure by which many weakly entangled qubit pairs are distilled into a
smaller number of more strongly entangled qubit pairs by means of local
operations and classical communication.

Alice and Bob's two-qubit state Eq.~(\ref{theform}) can be
distilled successfully using local operations and classical
communication as long as they satisfy the Peres--Horodecki partial
transposition criterion \cite{HOR-PER}: A two-qubit state
$\varrho$ is quantum distillable if and only if it is a
\emph{non-positive partial transposed} (NPPT) state. A state
$\varrho$ is NPPT if $\varrho^{T_B}\not\ge 0$ so that it has at
least one negative eigenvalue. Here, $\varrho^{T_B}$ denotes the
transposition with respect to Bob's basis only. The partial
transpose of each of our Bell states gives,
\begin{eqnarray}
\ket{z_{kl}}\bra{z_{kl}} &\longrightarrow&
\frac{1}{2}-\ket{z_{k+1\; l+1}}\bra{z_{k+1\; l+1}}.
\end{eqnarray}

Applying the Peres--Horodecki criterion to our Bell diagonal
mixture, we find that the state Eq.~(\ref{theform}) is quantum
distillable provided that
\begin{eqnarray}\label{qd}
\max_{ab} p_{ab} &>& \frac{1}{2}.
\end{eqnarray}

\section{Advantage Distillation}

Instead of manipulating their qubits in QED, Alice and Bob can
process the raw key sequence they have established in the protocol
in order to obtain a more secure key sequence. One such procedure
is known as \emph{advantage distillation} (AD).

In the AD protocol, Alice and Bob divide their raw key sequence
into blocks of length $L$. For each block, Alice generates a
random bit and adds this, modulo $2$, to each bit of the block.
She then sends this processed block to Bob via a public channel.
After receiving the block, Bob subtracts his corresponding block
from it (modulo $2$). If all the bit values are the same, it is a
deemed a good block. Otherwise it is a bad block. Bob then informs
Alice whether the block he received was good or bad. If it is a
good block, Alice will record the random bit she initially
generated into her distilled bit sequence while Bob enters into
his distilled sequence the common bit value he found after
subtraction. If it is a bad block, they will both reject the bits
and it plays no further part in the distillation procedure.

Now for a good block, two cases can occur:
\begin{description}
    \item[(I)] Alice and Bob's distilled bits are the same;
    \item[(II)] Alice and Bob's distilled bits are different.
\end{description}
Case (I) occurs when Alice and Bob started out with an identical
raw block (i.e.\ their length $L$ blocks are perfectly
correlated). On the other hand, Case (II) occurs when Alice and
Bob start out with raw $L$-blocks that are anti-correlated with
each other.

Now, for large $L$, there will be approximately $\frac{L}{3}$ bits
in the good block that result from Alice and Bob's $z$ basis
measurement. For these, $p_0$ is the probability that Alice and
Bob obtain correlated results while $p_1$ is the probability that
they obtain anti-correlated results.  Similarly, $\frac{L}{3}$
bits will result from $x$ ($y$) basis measurement
--- $q_0$ ($r_0$) is the probability that Alice and Bob obtain
correlated results while $q_1$ ($r_1$) is the probability that
they obtain anti-correlated results. Thus for a good block, Case
(I) occurs with probability $\frac{p_0^{L/3} q_0^{L/3}
r_1^{L/3}}{p_0^{L/3} q_0^{L/3}
r_1^{L/3}+p_1^{L/3}q_1^{L/3}r_0^{L/3}}$ while Case (II) occurs
with probability $\frac{p_1^{L/3} q_1^{L/3} r_0^{L/3}}{p_0^{L/3}
q_0^{L/3} r_1^{L/3}+p_1^{L/3}q_1^{L/3}r_0^{L/3}}$ (remember that
for the $y$ basis, Alice and Bob generate their raw key from
anti-correlation, which corresponds to probability $r_1$). The
error rate for Alice and Bob (the proportion of Case (II) blocks)
is given by
\begin{eqnarray}
E_{AB} &=& \frac{p_1^{L/3} q_1^{L/3} r_0^{L/3}}{p_0^{L/3}
q_0^{L/3} r_1^{L/3}+p_1^{L/3}q_1^{L/3}r_0^{L/3}},
\end{eqnarray}
which for $L\gg 1$, $p_1<p_0$ and $q_1<q_0$ (since $p_{00} > \frac{1}{2}$) is approximately
\begin{eqnarray}
E_{AB} &\approx& \left(\frac{p_1 q_1 r_0}{p_0 q_0
r_1}\right)^{L/3}.
\end{eqnarray}

Eve is able to intercept the processed blocks that Alice sends to
Bob via the classical channel. From their public communication, she
will also be able to know which of the blocks are accepted or rejected. For
the good blocks, she has to deduce the distilled bit for each block.
To do this, she can either resort to incoherent or coherent measurements on
her ancillas.

\subsection{Incoherent Attack on Advantage Distillation}

In the incoherent attack, Eve performs a square root measurement to distinguish
her ancillas one by one and, from her results, tries to deduce what Alice and Bob measured for each entry in an $L$-block. She then subtracts Alice's transmitted
block from her own corresponding block, as Bob does. Typically, Eve's block
will be inhomogeneous after subtraction so she decides by majority voting which
bit value to assign to a particular block -- she bets on the value which
occurs most frequently in her block, and if there are the same number of $0$s
as $1$s, she picks one of them at random.

Consider Case (I) blocks, i.e. Alice and Bob start out with
correlated raw blocks. From her square-root measurement, Eve
guesses each entry in the block correctly with the following
probabilities:
\begin{itemize}
    \item $\eta_0^{x}$ if Alice and Bob measured in the $x$ basis;
    \item $\eta_1^{y}$ if they measured in the $y$ basis;
    \item $\eta_0^{z}$ if they measured in the $z$-basis.
\end{itemize}
She guesses an entry incorrectly with probabilities
\begin{itemize}
    \item $1-\eta_0^{x}$ if Alice and Bob measured in the $x$ basis;
    \item $1-\eta_1^{y}$ if they measured in the $y$ basis;
    \item $1-\eta_0^{z}$ if they measured in the $z$-basis.
\end{itemize}

Because Eve applies majority voting, she makes errors whenever she
guesses more than half of the entries in a block wrongly. If the
same number of $0$s and $1$s appear in her guesses, she picks one
of them at random and makes errors half of the time. We can thus
compute Eve's error rate:
\begin{eqnarray}\label{whaoC}
E^{(I)}_{BE} &=& \Sum{\sum_i
e_i>\frac{L}{2}}{}\ch{\frac{L}{3}}{e_x} (1-\eta^{x}_0)^{e_x}
(\eta^{x}_0)^{\frac{L}{3}-e_x}  \nonumber\\
&&\times \ch{\frac{L}{3}}{e_y} (1-\eta^{y}_1)^{e_y}
(\eta^{y}_1)^{\frac{L}{3}-e_y}\nonumber\\
&&\times\ch{\frac{L}{3}}{e_z} (1-\eta^{z}_0)^{e_z}
(\eta^{z}_0)^{\frac{L}{3}-e_z}
\nonumber\\
&&+\frac{1}{2}\Sum{\sum_i e_i=\frac{L}{2}}{}\ch{\frac{L}{3}}{e_x}
(1-\eta^{x}_0)^{e_x} (\eta^{x}_0)^{\frac{L}{3}-e_x}  \nonumber\\
&&\times \ch{\frac{L}{3}}{e_y}
(1-\eta^{y}_1)^{e_y} (\eta^{y}_1)^{\frac{L}{3}-e_y}\nonumber\\
&&\times \ch{\frac{L}{3}}{e_z}
(1-\eta^{z}_0)^{e_z} (\eta^{z}_0)^{\frac{L}{3}-e_z},
\end{eqnarray}
where $e_i$ is the number of errors made in the $i^{th}$ basis.
The second summation arises from the situation when Eve has to assign $0$
or $1$ at random to the block because the number of $0$s and $1$s in the block
are equal.

For $L\gg 1$, we can lower bound the summations in Eq.~(\ref{whaoC}) by
approximating them with the main contributing terms, i.e., terms for which the
binomial factor $\ch{\frac{L}{3}}{e_m}, (m=x,y,z)$ has its peak:
\begin{eqnarray}
E^{(I)}_{BE} &\sim& \ch{\frac{L}{3}}{\frac{L}{6}}
(1-\eta^{x}_0)^{\frac{L}{6}} (\eta_0^{x})^{\frac{L}{6}}
\nonumber\\
&&\times\ch{\frac{L}{3}}{\frac{L}{6}}
(1-\eta^{y}_1)^{\frac{L}{6}} (\eta_1^{y})^{\frac{L}{6}}\nonumber\\
&&\times\ch{\frac{L}{3}}{\frac{L}{6}}
(1-\eta^{z}_0)^{\frac{L}{6}} (\eta_0^{z})^{\frac{L}{6}}.
\end{eqnarray}
By applying Stirling's approximation we have
\begin{equation}
E^{(I)}_{BE} \sim 2^L \left(\eta_0^{x}\eta_1^{y}\eta_0^{z}
(1-\eta_0^{x})(1-\eta_1^{y})(1-\eta_0^{z})\right)^{\frac{L}{6}}.
\end{equation}

Similarly for Case (II) blocks in which Alice and Bob start out with
anti-correlated raw blocks, we can obtain the error rate for Eve:
\begin{equation}
E^{(II)}_{BE} \sim 2^L \left(\eta_1^{x}\eta_0^{y}\eta_1^{z}
(1-\eta_1^{x})(1-\eta_0^{y})(1-\eta_1^{z})\right)^{\frac{L}{6}}.
\end{equation}

Finally, the total error rate for Eve is given by
\begin{eqnarray}
E_{BE} &\sim&
\frac{p_0^{\frac{L}{3}}q_0^{\frac{L}{3}}r_1^{\frac{L}{3}}}{p_0^{\frac{L}{3}}
q_0^{\frac{L}{3}}r_1^{\frac{L}{3}}+p_1^{\frac{L}{3}}q_1^{\frac{L}{3}}r_0^{\frac{L}{3}}}
E^{(I)}_{BE}  \nonumber\\
&& +\frac{p_1^{\frac{L}{3}}
q_1^{\frac{L}{3}}r_0^{\frac{L}{3}}}{p_0^{\frac{L}{3}}
q_0^{\frac{L}{3}}r_1^{\frac{L}{3}}+p_1^{\frac{L}{3}}q_1^{\frac{L}{3}}r_0^{\frac{L}{3}}}
E^{(II)}_{BE}
\end{eqnarray}

Since the coefficient in front of $E^{(I)}_{BE}$ goes to 1 while the coefficient in
front of $E^{(II)}_{BE}$ goes to 0, we are left with

\begin{equation}
E_{BE} \approx  2^L  \left(\eta_0^{x}\eta_1^{y}\eta_0^{z}
(1-\eta_0^{x})(1-\eta_1^{y})(1-\eta_0^{z})\right)^{\frac{L}{6}}.
\end{equation}

By comparing the error rates \cite{Maurer}, we can obtain the condition for AD
to be successful under an incoherent attack:
\begin{equation}
\lim_{L\rightarrow\infty}\frac{E_{AB}}{E_{BE}} < 1
\label{eq:CKcond}
\end{equation}
which reduces to
\begin{equation}\label{cadIncoh}
\frac{p_1}{p_0}\frac{q_1}{q_0}\frac{r_0}{r_1} <
8\sqrt{\eta_0^{x}\eta_1^{y}\eta_0^{z}
(1-\eta_0^{x})(1-\eta_1^{y})(1-\eta_0^{z})}.
\end{equation}

For the special case of Werner states
($p_{01}=p_{10}=p_{11}=\frac{1-p_{00}}{3}$, so that $p_0=q_0=r_1$,
$p_1=q_1=r_0$ and $\eta_0^x=\eta_1^y=\eta_0^z$), we find that
Eq.~(\ref{cadIncoh}) reduces to
\begin{eqnarray}
\frac{p_1}{p_0} &<& 2\sqrt{\eta_0^z (1-\eta_0^z)}.
\end{eqnarray}
A similar result was obtained by Bru$\ss$ et al.\ \cite{cadIN}.

\subsection{Coherent Attack on Advantage Distillation}

We consider a particularly simple scheme of coherent attack that
is similar to that presented in \cite{K}. Eve's strategy is as
follows.

For each good block, Eve has a corresponding set of ancilla states and rather than measuring her ancillas one-by-one (an incoherent attack), she performs a joint
measurement on \emph{all} $L$ of them to acquire knowledge about the value that
Alice assigned to the block. By also making use of the classical information
that is exchanged between Alice and Bob during the distillation process, Eve
can learn a lot more than if she were to measure her ancillas one by one.

Consider first a Case (I) block.
As an example, suppose that Alice and Bob start out with the same
block $01001$ for $L=5$, and Alice's random bit is $1$. After
addition (modulo $2$), she sends the processed block $10110$ to
Bob via the public channel which Eve is able to intercept. Eve can
also project her block of ancilla states into the orthogonal
subspace corresponding to Alice and Bob having a correlated or
anti-correlated block. Doing this, she can know that Alice and Bob
started out with the same raw blocks (i.e.\ Case (I) blocks). From
this, Eve can then deduce the following possibilities:
\begin{enumerate}
\item If Alice's random bit is `$0$', Alice and Bob must have started out with
  raw blocks $10110$. If Alice and Bob had measured in the bases $x,y,y,z,x$ for the respective entries in the block, the \
  ancilla state that she holds will be
  $\ket{f^{x}_{11}}\ket{f^{y}_{01}}\ket{f^{y}_{10}}
  \ket{f^{z}_{11}}\ket{f^{x}_{00}}$.
\item If Alice's random bit is `$1$', Alice and Bob must have started out with
  raw blocks $01001$. If $x,y,y,z,x$ is the order of basis measurements for the entries,
  the ancilla state that she holds will then be
  $\ket{f^{(x)}_{00}}\ket{f^{y}_{10}}\ket{f^{y}_{01}}
  \ket{f^{z}_{00}}\ket{f^{x}_{11}}$.
\end{enumerate}
The mutual inner product between the two ancilla states is
$(\lambda^{x}_0)^{n_x}
(\lambda^{y}_1)^{n_y}(\lambda^{z}_0)^{n_z}$, where $n_a$ is the
number of times the basis $a$ was measured. The optimal
measurement to distinguish these two states is again the square
root measurement. In general, for each Case (I) block of length
$L$, Eve needs to distinguish just $2$ possible $L$-ancilla states
with mutual inner product $(\lambda^{x}_0)^{n_x}
(\lambda^{y}_1)^{n_y} (\lambda^{z}_0)^{n_z}$.

Now, for large $L$, we have $n_x, n_y, n_z \approx \frac{L}{3}$. Eve's
probability of correctly inferring a particular $L$-ancilla state is
given by
\begin{equation}
\frac{1}{2}\left(
1+\sqrt{1-(\lambda^{x}_0\lambda^{y}_1\lambda^{z}_0)^{\frac{2L}{3}}}\right)
\approx 1-\frac{1}{4}
(\lambda^{x}_0\lambda^{y}_1\lambda^{z}_0)^{\frac{2L}{3}}.
\end{equation}
Her error rate for Case (I) blocks is thus
\begin{equation}
E_{BE}^{(I)}\approx
\frac{1}{4}(\lambda^{x}_0\lambda^{y}_1\lambda^{z}_0)^{\frac{2L}{3}}.
\end{equation}

Similarly when we consider Case (II) blocks,
Eve's error rate is
\begin{equation}
E_{BE}^{(II)}\approx \frac{1}{4}
(\lambda^{x}_1\lambda^{y}_0\lambda^{z}_1)^{\frac{2L}{3}}.
\end{equation}

Eve's \emph{total} error rate is thus
\begin{eqnarray}
E_{BE} &=&
\frac{p_0^{\frac{L}{3}}q_0^{\frac{L}{3}}r_1^{\frac{L}{3}}}{p_0^{\frac{L}{3}}
q_0^{\frac{L}{3}}r_1^{\frac{L}{3}}+p_1^{\frac{L}{3}}q_1^{\frac{L}{3}}r_0^{\frac{L}{3}}}
E_{BE}^{(I)}  \nonumber\\
&&
+\frac{p_1^{\frac{L}{3}}q_1^{\frac{L}{3}}r_0^{\frac{L}{3}}}{p_0^{\frac{L}{3}}
q_0^{\frac{L}{3}}r_1^{\frac{L}{3}}+p_1^{\frac{L}{3}}q_1^{\frac{L}{3}}r_0^{\frac{L}{3}}}E_{BE}^{(II)}  \nonumber\\
&\approx& \frac{1}{4}
(\lambda^{x}_0\lambda^{y}_1\lambda^{z}_0)^{\frac{2L}{3}}
\end{eqnarray}
since once again, the coefficient in front of $E^{(I)}_{BE}$ goes to 1 while the coefficient in front of $E^{(II)}_{BE}$ goes to 0.

Finally by comparing error rates (Eq.~(\ref{eq:CKcond})), we obtain the
condition for AD to be possible under a coherent attack by Eve:
\begin{equation}\label{cadCo}
\frac{p_1}{p_0}\frac{q_1}{q_0}\frac{r_0}{r_1} <
\left(\lambda^{x}_0\lambda^{y}_1\lambda^{z}_0\right)^2.
\end{equation}

\section{Discussion}

We now analyze the above results. A Bell diagonal density matrix
is characterized by four real parameters and a normalization
condition so we can parameterize such a state by the probability
$p_{00}$ (the amount of the state $|z_{00}\rangle$ in the Bell
mixture) and two angles $\theta,\phi$ characterizing the remaining
three probabilities $p_{01},p_{10},p_{11}$:
\begin{eqnarray}
p_{01} &=& (1-p_{00})\cos^2\theta\cos^2\phi\nonumber\\
p_{10} &=& (1-p_{00})\sin^2\theta\cos^2\phi\nonumber\\
p_{11} &=& (1-p_{00})\sin^2\phi.
\end{eqnarray}
This means that for a fixed $p_{00}$, all the quantities such as
$I_{AB},I_{BE},E_{AB},E_{BE}$ for incoherent and coherent attacks
are two-argument functions.

\begin{figure*}[!h]
\includegraphics[width=\textwidth]{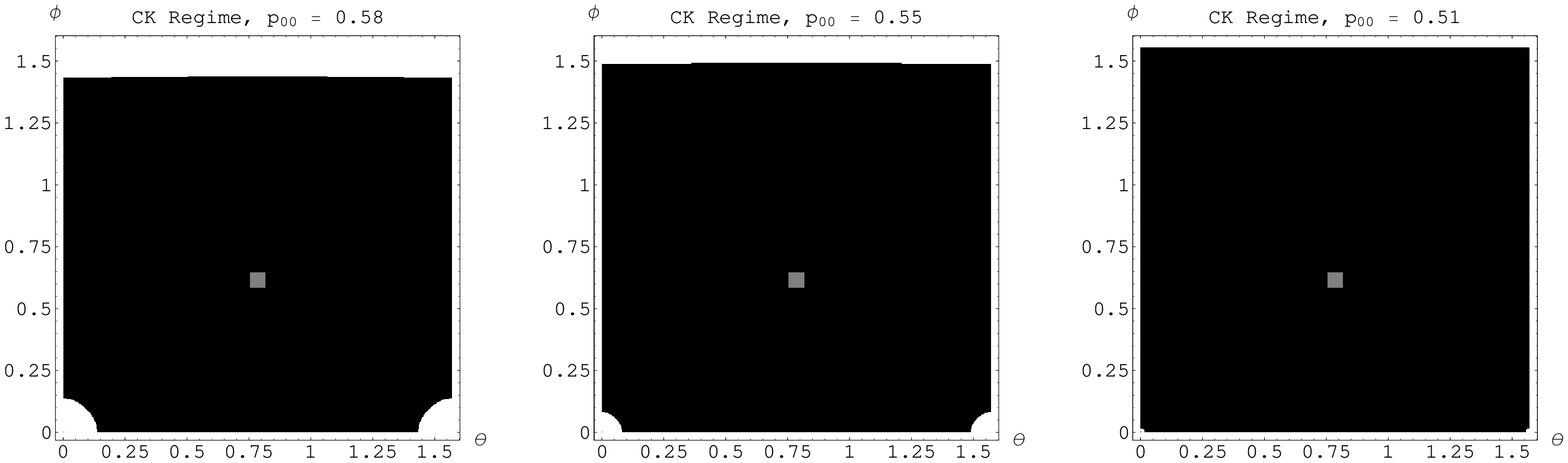}
\caption{Comparison of secure regions for the protocol for
different values of $p_{00}$ under an incoherent attack. White
regions in the plot represent states that are secure against
incoherent attacks by Eve in the scenario when Alice and Bob do
not attempt AD or QED (CK regime). When $p_{00}\rightarrow
\frac{1}{2}$ the white areas disappear with the exception of the
certain points that never become black. These points correspond to
the states $p_{00}|z_{00}\rangle\langle z_{00}|+
p_{01}|z_{01}\rangle\langle z_{01}|$ ($\theta=0$, $\phi=0$),
$p_{00}|z_{00}\rangle\langle z_{00}|+ p_{10}|z_{10}\rangle\langle
z_{10}|$ ($\theta=\frac{\pi}{2}$, $\phi=0$) and
$p_{00}|z_{00}\rangle\langle z_{00}|+ p_{11}|z_{11}\rangle\langle
z_{11}|$ ($\phi=\frac{\pi}{2}$). These states are resistant to any
incoherent attack. As reference, the grey areas (exaggerated in
the figure) indicate Werner states.} \label{fig:fig1}
\end{figure*}

First, for each $p_{00}$ we can plot a region characterizing all
the Bell diagonal states which lead to secure \emph{raw} keys. As
long as $p_{00}$ is greater than around $0.765$ all corresponding
states are secure. Below this, fewer and fewer states are secure
(white regions in Fig.~\ref{fig:fig1}) until, for
$p_{00}=\frac{1}{2}$, the Bell diagonal mixture becomes separable
and no secret bits can be obtained. Even then we can still
identify certain states that are resistant against incoherent
eavesdropping as long as $p_{00}$ is greater than half. These are
states of the form
$p_{00}\ket{z_{00}}\bra{z_{00}}+p_{01}\ket{z_{01}}\bra{z_{01}}$,
$p_{00}\ket{z_{00}}\bra{z_{00}}+p_{10}\ket{z_{10}}\bra{z_{10}}$
and
$p_{00}\ket{z_{00}}\bra{z_{00}}+p_{11}\ket{z_{11}}\bra{z_{11}}$.
It is interesting to note that this threshold of $0.765$, below
which it is no longer possible to generate secure keys for every
state, is the same threshold as that for the Werner state ---
this means that the Werner state will be the first state to become
insecure as the $p_{00}$ threshold is exceeded.

Second, using Eq.~(\ref{cadIncoh}) we verified the results
presented in~\cite{ACIN}, namely that QED is equivalent to AD if
Eve can only perform incoherent attacks. In other words, as long
as $p_{00}$ is greater than $\frac{1}{2}$, Alice and Bob do not
need QED because AD works equally well and does not require
collective operations on qubits, which are difficult to realize
experimentally.

\begin{figure*}[!h]
\includegraphics[width=\textwidth]{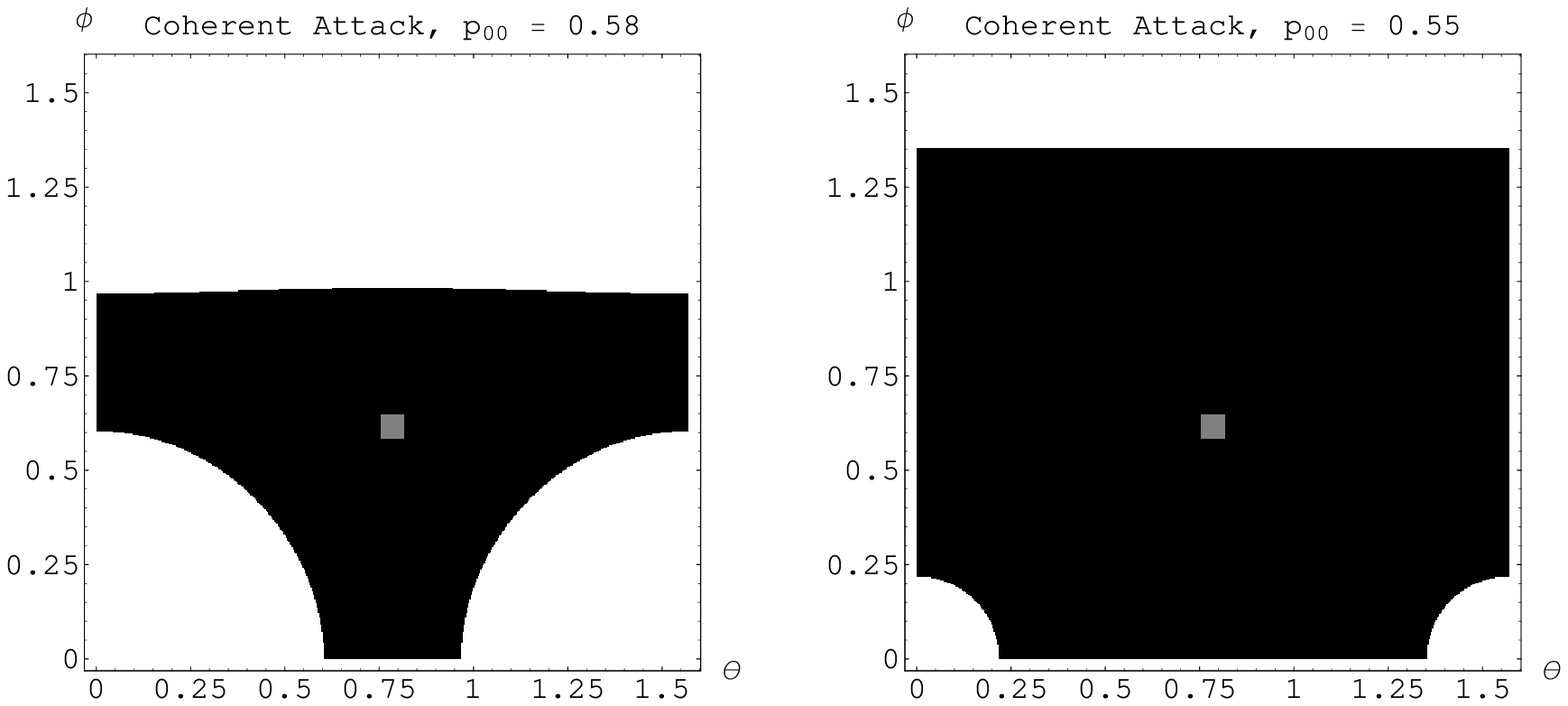}
\caption{Comparison of secure regions in advantage distillation
for different values of $p_{00}$ under a coherent attack. White
regions in the plot represent states
  that are secure against coherent attacks by Eve in the scenario when Alice
  and Bob perform AD. Black regions correspond to states for which AD fails
  under coherent attack. As the state becomes more mixed
  ($p_{00}\rightarrow\frac{1}{2}$), the white areas disappear with the
  exception of the certain points that never become black. As with the CK
  regime for $p_{00}\rightarrow \frac{1}{2}$, the surviving states are $p_{00}|z_{00}\rangle \langle z_{00}| + p_{01}|z_{01}\rangle\langle z_{01}|$ , $p_{00} |z_{00}\rangle \langle z_{00}|+p_{10}|z_{10}\rangle \langle z_{10}|$ and $p_{00}|z_{00}\rangle \langle z_{00}| +p_{11}|z_{11}\rangle\langle z_{11}|$.  In comparison, with only incoherent attacks all states with $p_{00}>\frac{1}{2}$ are secure. As reference, the grey areas (exaggerated in the figure) refer to Werner states.}
\label{fig:fig2}
\end{figure*}

However, if Eve is capable of carrying out a coherent attack, QED
is much more powerful than AD (Fig.~\ref{fig:fig2}). We see that
as $p_{00}\rightarrow\frac{1}{2}$, more states fall into the black
regions where AD fails and only QED is possible. As before, the
same states that are resistant to incoherent attack in the CK
regime are resistant to the above coherent attack on AD.

\section{Conclusion}

We have generalized the tomographic QKD scheme to Bell diagonal states and
analyzed its resistance to various eavesdropping attacks, both in the CK
regime and when Alice and Bob perform advantage distillation. We have shown the
inequivalence of advantage distillation and entanglement distillation in the
presence of coherent measurement by a potential eavesdropper. It still remains
to be seen whether Eve can further increase her information gain by entangling
more than one pair of Alice and Bob's qubits with her ancilla.

DKLO is supported by the Cambridge-MIT Institute project on
quantum information and Sidney Sussex College Cambridge, and
acknowledges EU grants RESQ (IST-2001-37559) and TOPQIP
(IST-2001-39215). DK, LCK and AG wish to acknowledge support from
A*STAR Grant R-144-000-071-305. DK wishes to acknowledge NUS Grant
R-144-000-089-112. DK and AG also wish to thank Antonio Ac\'{\i}n
for valuable discussions.



\begin{thebibliography}{10}
\bibitem{p00} Here we are assuming that one of the $p_{ab}$'s must
be greater than $\frac{1}{2}$ as otherwise the state is separable
and thus insecure. Without loss of generality, we can always
assume that $p_{00}>\frac{1}{2}$ with the corresponding state $\ket{z_{00}}$ being Alice and Bob's ``ideal'' state.
\end{thebibliography}
\end{document}